 \documentclass[10pt,twocolumn]{article}

 \pdfoutput=1

\usepackage[round]{natbib}   

\usepackage[english]{babel}

\usepackage{graphicx}
\usepackage{amsmath}

\usepackage{enumitem}

\usepackage{float}
\usepackage{ifthen}
\usepackage{hyperref}

\usepackage{titlesec}

\titlespacing*{\section}{0pt}{1.1\baselineskip}{\baselineskip}

\begin{document}



\twocolumn[
\begin{center}
{\bf \Large 
Comments on ``Isentropic Analysis of a Simulated Hurricane''}\\
\vspace*{3mm}
{\Large by Pascal Marquet}. \\
\vspace*{3mm}
{\large  M\'et\'eo-France, CNRM/GMAP-CNRS UMR-3589.
 Toulouse. France.}
\\ \vspace*{2mm}
{\large  \it E-mail: pascal.marquet@meteo.fr}
\\ \vspace*{2mm}
{\large     Submitted to the}
{\large \it Journal of Atmospheric Science}
{\large (13 July 2016).} \\
\vspace*{1mm}
\end{center}
]


 \section{\large Introduction} 
\vspace*{-4 mm}

In a recent paper, \citet[hereafter MPZ]{Mrowiec_al_2016} investigated the thermodynamic properties of a three-dimensional hurricane simulation.
The paper MPZ focused on isentropic analysis based on the conditional averaging of the mass transport with respect to the equivalent potential temperature $\theta_e$, with the underlying assumption that $\theta_e$ defined in \citet[hereafter E94]{Emanuel94} is a logarithmic measurement of the moist-air entropy.
It was also assumed that isentropic surfaces are represented by constant values of $\theta_e$.

Many other equivalent potential temperatures exist in the literature however and it is shown in this comment that the way in which the moist entropy $s$ and these equivalent potential temperatures are defined may lead to opposite results in studies of isentropic processes in hurricanes.
It is shown that the more the total water varies with space the more the isentropes differ.

The paper is organized as follows.
Different potential temperatures are recalled in section~\ref{section_theta} and the associated moist-air entropy is presented in section~\ref{section_s}.
A data-set derived from a simulation of the hurricane DUMILE is presented in section~\ref{section_Dumile}.
Section~\ref{section_impact_theta} provides numerical evaluations for $\theta_e$, $\theta_{es}$ (the saturated version of $\theta_e$) and $\theta_s$ defined by \citet[hereafter M11]{Marquet11}, where $\theta_s$ is computed by applying the third law of thermodynamic, a process improved by \citet{Marquet15,Marquet16}.
Differences in the associated moist-air entropies are described in section~\ref{section_impact_s}, together with an evaluation of the heat input computed for a Carnot cycle in the so-called temperature-entropy diagram.
A conclusion is presented in section~\ref{conclusions}.

 \section{\large The moist-air potential temperatures} 
\label{section_theta}
\vspace*{-4 mm}

The equivalent potential temperature $\theta_e$ defined by Eq.~(4.5.11) in \citet{Emanuel94} can be written as
\vspace*{-1mm}
\begin{align}
 \!\!\!
  {\theta}_{e/E94} 
   & = \: T \; \left( \frac{p_0}{p_d}\right)^{\!\!R_d/c^{\ast}_{pl} } \:
    \exp\! \left( \frac{L_v \: r_v}{c^{\ast}_{pl}  \: T} \right) \:
    {(H_l)}^{- \, R_v \, r_v / c^{\ast}_{pl}}
\label{eq_thetae_E94} \: ,
\end{align}
where $T$ is the temperature, $p_0$ is the standard pressure, $p_d$ is the dry-air pressure, and $R_d$ and $R_v$ are the gas constant of water vapor and dry air, respectively.
The specific heat $c^{\ast}_{pl} = c_{pd} + r_t \: c_l$ depends on the values for dry air ($c_{pd}$) and liquid water ($c_l$).
The mixing ratios $r_v$ and $r_t$ represent water vapor and total water, respectively.
$L_v$ is the latent heat of vaporization.
The relative humidity with respect to liquid water $H_l = e/e_{sw}$ is the ratio of water vapor pressure ($e$) over the saturated value ($e_{sw}$).

The saturated equivalent potential temperature studied in \citet[hereafter E86]{Emanuel_86} will be written as
\vspace*{-1mm}
\begin{align}
  \! \! {\theta}_{es/E86} 
   & = \: T \: \left( \frac{p_0}{p}\right)^{\!\!R_d/c^{\ast}_{pl} } \:
    \exp\! \left( \frac{L_v \: r_{sw}}{c^{\ast}_{pl}  \: T} \right)
\label{eq_thetaes} \: ,
\end{align}
where $p$ is the total pressure and $r_{sw}(T,p)$ is the saturation mixing ratio at temperature $T$ and pressure $p$.

The equivalent potential temperature studied in \citet[hereafter B73]{Betts73} can be written as
\vspace*{-1mm}
\begin{align}
  {\theta}_{e/B73} 
   & = \: \theta \; \:
    \exp\! \left( \frac{L_v \: q_v}{c_{pd}  \: T} \right)
\label{eq_thetae_B73} \: ,
\end{align}
where the dry-air potential temperature is $\theta = T \: (p_0/p)^{\kappa}$ with $\kappa = R_d/c_{pd} \approx 0.286$.

The equivalent potential temperature $\theta_e$ studied in MPZ can be written as
\vspace*{-1mm}
\begin{align}
  \! \! \! \! {\theta}_{e/MPZ} 
   & = \: T \; \left( \frac{p_0}{p}\right)^{\!\!R_d/c^{\ast}_{pl} } \:
    \exp\! \left( \frac{L_v \: r_v}{c^{\ast}_{pl}  \: T} \right) \:
    {(H_l)}^{- \, R_v \, r_v / c^{\ast}_{pl}}
\label{eq_thetae_MPZ16} \: .
\end{align}
The difference between ${\theta}_{e/MPZ}$ and ${\theta}_{e/E94}$ given by Eq.~(\ref{eq_thetae_E94}) is $p_d$ replaced by $p$ in the Exner function.
The differences between ${\theta}_{e/MPZ}$ and  ${\theta}_{e/E86}$ given by Eq.~(\ref{eq_thetaes}) are $r_{sw}$ replaced by $r_v$ and  the additional term depending on ${H_l}$ included.
The differences between ${\theta}_{e/MPZ}$ and  ${\theta}_{e/B73}$ given by Eq.~(\ref{eq_thetae_B73}) are $c_{pd}$ replaced by $c^{\ast}_{pl}$, $q_v$ replaced by $r_v$ and the additional term depending on ${H_l}$ included.

The third-law based potential temperature defined in \citet{Marquet11,Marquet16} can be written as
\vspace*{-1mm}
\begin{align}
  {\theta}_{s/M11} 
   & = \: \theta \; \:
    \exp\! \left( - \: \frac{L_v \: q_l + L_s \: q_i}{c_{pd} \: T} \right)
       \:
    \exp\! \left(  \Lambda_r \: q_t  \right)
\nonumber \\
   &
        \left( \frac{T}{T_r}\right)^{\!\!\lambda \,q_t}
 \!  \! \left( \frac{p}{p_r}\right)^{\!\!-\kappa \,\delta \,q_t}
 \!  \! \left( \frac{r_r}{r_v} \right)^{\!\!\gamma\,q_t}
      \frac{(1\!+\!\eta\,r_v)^{\,\kappa \, (1+\,\delta \,q_t)}}
           {(1\!+\!\eta\,r_r)^{\,\kappa \,\delta \,q_t}}
\nonumber \\
   & \; {(H_l)}^{\, \gamma \, q_l} \;
     \; {(H_i)}^{\, \gamma \, q_i}
     \; {\left( \frac{T_l}{T} \right)}^{\! c_l\, q_l/c_{pd}}
     \; {\left( \frac{T_i}{T} \right)}^{\! c_i\, q_i/c_{pd}}
\label{eq_thetas} \: ,
\end{align}
where 
$\lambda = c_{pv}/c_{pd}-1 \approx 0.837$,
$\eta = R_v/R_d \approx 1.608$, 
$\delta = \eta - 1\approx 0.608$
and
$\gamma = R_v/c_{pd} \approx 0.46$.
The term $\Lambda_r = [ \: {(s_v)}_r - {(s_d)}_r \: ]/c_{pd} \approx 5.87$ depends on the third-law values for the reference entropies for water vapor and dry air, and $L_s$ is the latent heat of sublimation.
The specific contents $q_v$, $q_l$, $q_l$, $q_i$ and $q_t = q_v + q_l + q_i$ replace the mixing ratios involved in most of the previous formulations.

The four terms in the last line of Eq.~(\ref{eq_thetas}) derived in \citet{Marquet16} are improvements with respect to \citet{Marquet11}.
They take into account possible non-equilibrium processes such as under- or super-saturation with respect to liquid water ($H_l \neq 1$) or ice ($H_i \neq 1$), temperatures of rain $T_l$ or snow $T_i$ which may differ from those $T$ of dry air and water vapor.

The advantage of the term ${(H_l)}^{\, \gamma \, q_l}$ in Eq.~(\ref{eq_thetas}) compared with ${(H_l)}^{- \, R_v \, r_v / c^{\ast}_{pl}}$ in Eqs.~(\ref{eq_thetae_E94}) or (\ref{eq_thetae_MPZ16}) is that $q_l$ replaces $r_v$ in the exponent, making no impact in clear-air, under- or super-saturated moist regions (where $H_l \neq 1$ and $r_v$ may be large, but where $q_l=0$) and making lesser impact in cloud in under- or super-saturated regions (where $H_l \neq 1$ but where typically $q_l \ll r_v$).

The first and second order approximations of $\theta_s$ are derived in \citet{Marquet15}, leading to
\vspace*{-1mm}
\begin{align}
  {({\theta}_{s})}_1
   & \approx \: \theta \; \:
    \exp\! \left( - \: \frac{L_v \: q_l + L_s \: q_i}{c_{pd} \: T} \right)
       \:
    \exp\! \left(  \Lambda_r \: q_t  \right)
\label{eq_thetas1} \: , \\
  {({\theta}_{s})}_2
   & \approx \: {({\theta}_{s})}_1 \;
    \exp\! \left[ \: - \: \gamma \: (q_l+q_i) \right]
  \; {\left( \frac{r_v}{r_{\ast}} \right)}^{\! - \: \gamma \: q_t}
\label{eq_thetas2} \: ,
\end{align} 
where $r_{\ast} \approx 0.0124$~kg~kg${}^{-1}$.
Both ${({\theta}_{s})}_1$ and ${({\theta}_{s})}_2$ must be multiplied by the last line of Eq.~(\ref{eq_thetas}) if non-equilibrium processes are to be described.

 \section{\large The moist-air entropies} 
\label{section_s}
\vspace*{-4 mm}

The moist-air entropy is computed in M11 from the third-law of thermodynamics.
It can be written as
\vspace*{-1mm}
\begin{align}
  s(\theta_{s/M11})
   & = \; \:
  s_{\rm ref} \: + \: c_{pd} \: \ln(\theta_s)
\label{eq_s_thetas_M11} \: ,
 \\
  s(\theta_{s/M11})/q_d
   & = \; \:
  [ \: s_{\rm ref}/q_d \: ] \: + \: [ \: c_{pd}/q_d \: ] \: \ln(\theta_s)
\label{eq_s_thetas_M11_oqd} \: ,
\end{align} 
where both $s_{\rm ref} \approx 1139$~J~K${}^{-1}$~kg${}^{-1}$ and $c_{pd}$ are constant, making $\theta_s$ a true equivalent of the specific moist-air entropy $s$.
The second formulation $s(\theta_{s/M11})/q_d$ is expressed ``per unit of dry air'', in order to be better compared with the entropies computed in other studies such as E94 or MPZ.

Other definitions of ``moist-air entropy'' are derived with either $s_{\rm ref}$ or $c_{pd}$ (often both of them) depending on the total-water mixing ratio $r_t$.
This is true in Eq.~(4.5.10) in E94, which can be written as
\vspace*{-1mm}
\begin{align}
  s(\theta_{e/E94}) / q_d
   & = \; [ \: - \: R_d \ln(p_0)  \: ] \: + \: c^{\ast}_{pl} \: \ln(\theta_{e/E94})
\label{eq_s_thetae_E94} \: ,
\end{align}
where $p_0$ is a constant standard value.
The division of $s$ by $q_d$ means that the entropy in Eq.~(\ref{eq_s_thetae_E94}) is expressed ``per unit mass of dry air''.
The reference values of entropies disagree in E94 with the third law, the consequence being that several terms are missing or are set to zero in Eq.~(\ref{eq_s_thetae_E94}).
These missing terms may impact the specific entropy if $q_t$ varies in space or time, since these missing terms must be multiplied by $q_d = 1 - q_t$ to compute $s$ from $s/q_d$ given by Eq.~(\ref{eq_s_thetae_E94}).
Moreover, since $c^{\ast}_{pl} = c_{pd} + r_t \: c_l$ depends on $r_t$, changes in $s$ and $s/q_d$ cannot be represented by $\theta_{e/E94}$ in Eq.~(\ref{eq_s_thetae_E94}) for varying values of $r_t$.
This prevents $\theta_{e/E94}$ from being a true equivalent of the specific moist-air entropy $s$ for hurricanes, where properties of saturated regions (large values of $r_t$) are to be compared with non-saturated ones (small values of $r_t$).

Similarly, the moist-air entropy defined in section~3 in MPZ can be written as
\vspace*{-1mm}
\begin{align}
  s(\theta_{e/MPZ}) / q_d
   & = \;  [ \: - \: c^{\ast}_{pl} \: \ln(T_0) \: ] \: + \: c^{\ast}_{pl} \: \ln(\theta_{e/MPZ})
\label{eq_s_thetae_MPZ} \: ,
\end{align}
where $T_0$ is a constant standard value.
Again, it is an entropy expressed ``per unit mass of dry air'' and the specific heat $c^{\ast}_{pl}$ depends on varying values of $r_t$, twice preventing $\theta_{e/MPZ}$ from being a true equivalent of the specific moist entropy.

It is shown in \citet[section 5.3]{Marquet_Geleyn_2015} that the linear combination $s_a = (1-a)\:s(\theta_e) + a\:s(\theta_l)$ described in the Appendix~C of \citet{Pauluis_al_2010} can lead to the  third law value of entropy $s(\theta_{s/M1})$ if the weighting factor is set to the value $a \approx 0.356$.
Another value for $a$ would lead to a definition of the specific moist-air entropy $s_a$ which would disagree with the third law.

In order to better analyze the impact of $r_t$ on the term $c^{\ast}_{pl}$, and thus on the definition of the moist-air entropy, two kinds of ``saturated equivalent entropy'' are defined.
They are based on the definition of $\theta_{es/E86}$ given by Eq.~(\ref{eq_thetaes}), yielding 
\vspace*{-1mm}
\begin{align}
  s(\theta_{es/E86}) / q_d
   & = \: c^{\ast}_{pl} \: \ln(\theta_{es/E86})
\label{eq_s_theta_es_E86_cpast} \: , \\
  s(\theta_{es/E86}) / q_d
   & = \: c_{pd} \: \ln(\theta_{es/E86})
\label{eq_s_theta_es_E86_cpd} \: .
\end{align}

 \section{\large The data set for the hurricane DUMILE} 
\label{section_Dumile}
\vspace*{-4 mm}

\begin{figure}[hbt]
\centerline{\includegraphics[width=0.9\linewidth]{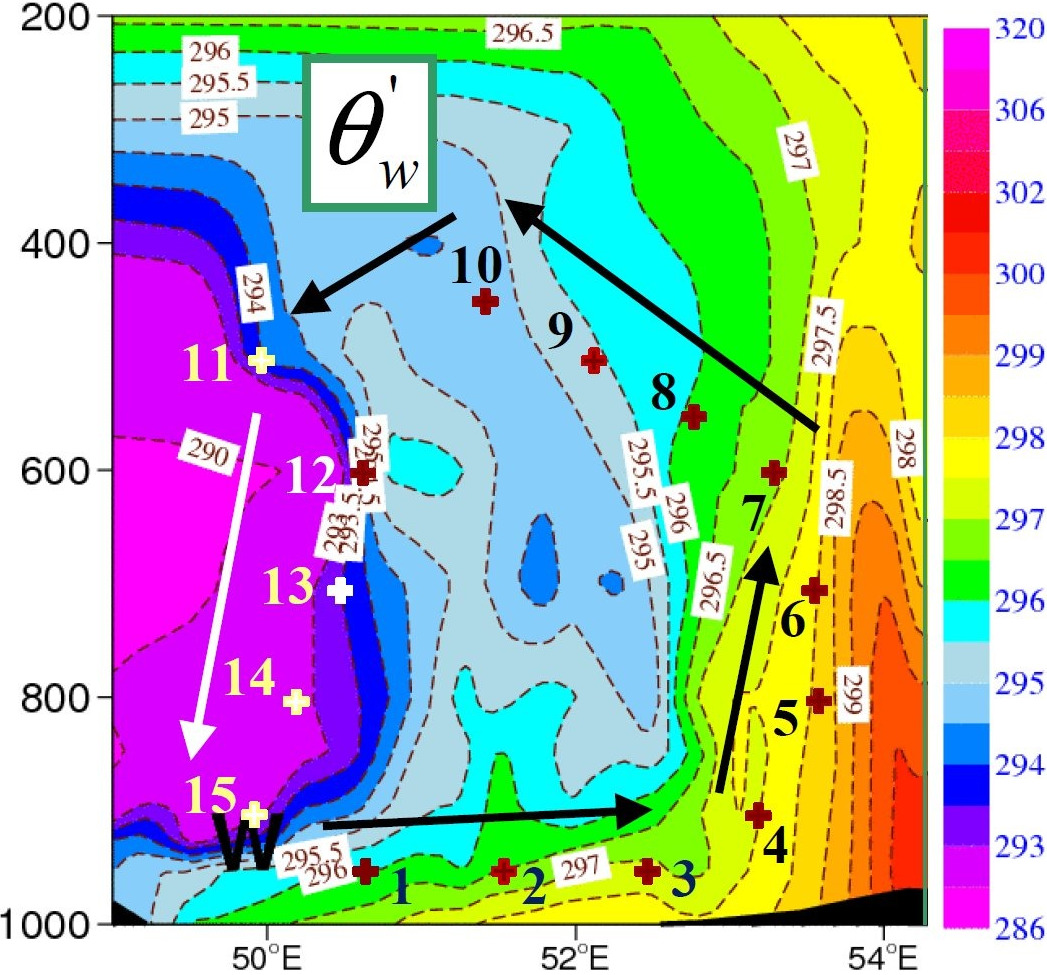}}
\vspace*{-3mm} 
\caption{\it A pressure-longitude cross section at $18.5$ south latitude for the hurricane Dumile and for the pseudo-adiabatic potential temperature $\theta'_w$ in K.
The $15$ points are used to build the Table~\ref{Table1} and to plot the Figs.~\ref{fig5}-\ref{fig6}.
} 
\label{fig1}
\end{figure}

\begin{figure}[hbt]
\centerline{\includegraphics[width=0.93\linewidth]{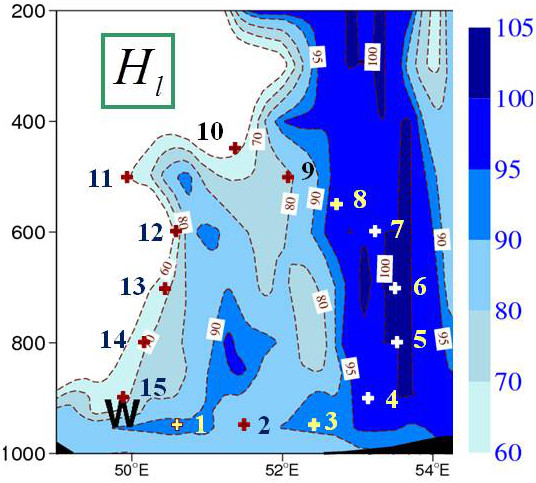}}
\vspace*{-3mm} 
\caption{\it The same as in Fig.~\ref{fig1}, but for the relative humidity $H_l$ in \%.
} 
\label{fig2}
\end{figure}

On January 3, 2013 at 00 UTC, the hurricane Dumile was located northwest of the R\'eunion island and east of Madagascar, near $18.5$ south latitude and $54.25$ east longitude.
Two cross sections are depicted in Figs.~\ref{fig1} and \ref{fig2} for the pseudo-adiabatic potential temperature $\theta'_w$ and the relative humidity $H_l$.
The use of $\theta'_w$ allows a clear, unambiguous definition of thermal properties, differing from the uncertain and multiple definitions of $\theta_e$ recalled in section~\ref{section_theta} which are to be compared in later sections.

The pressure is used as a vertical coordinate and the black regions close to $1000$~hPa represent the east coast of Madagascar on the left, the center of Dumile on the right.
The west-east cross sections are plotted for a $12$~hour forecast by employing the French model ALADIN, with a resolution of about $8$~km.

\begin{figure}[hbt]
\centerline{\includegraphics[width=0.9\linewidth]{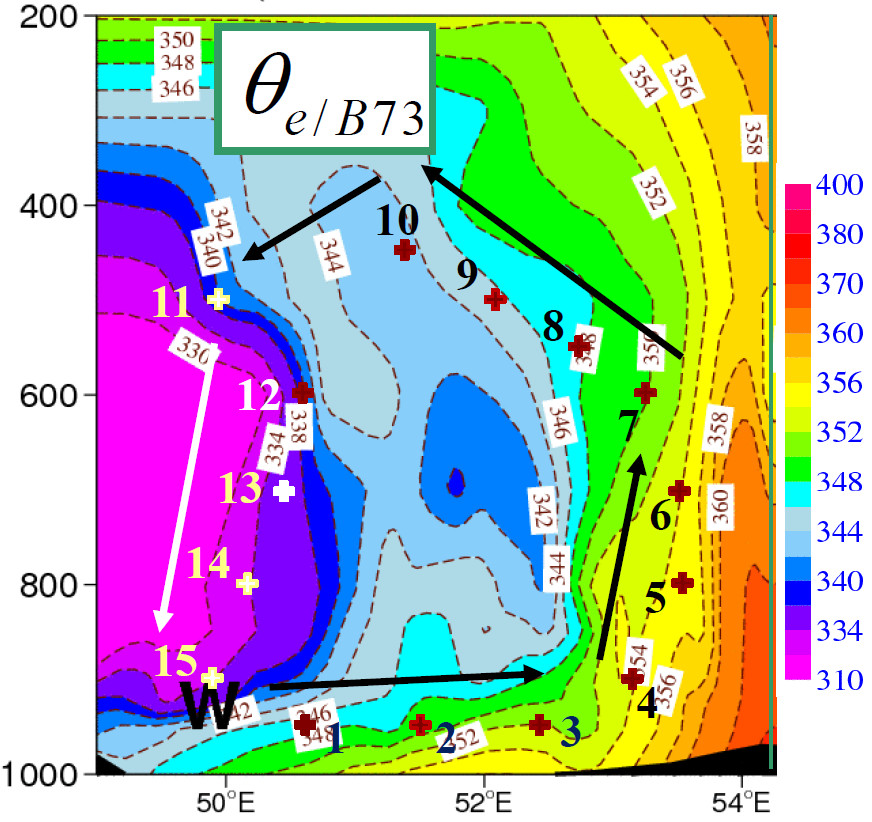}}
\vspace*{-3mm} 
\caption{\it The same as in Fig.~\ref{fig1}, but for the potential temperatures $\theta_{e/B73}$ in K computed by Eq.~(\ref{eq_thetae_B73}).
} 
\label{fig3}
\end{figure}

\begin{figure}[hbt]
\centerline{\includegraphics[width=0.9\linewidth]{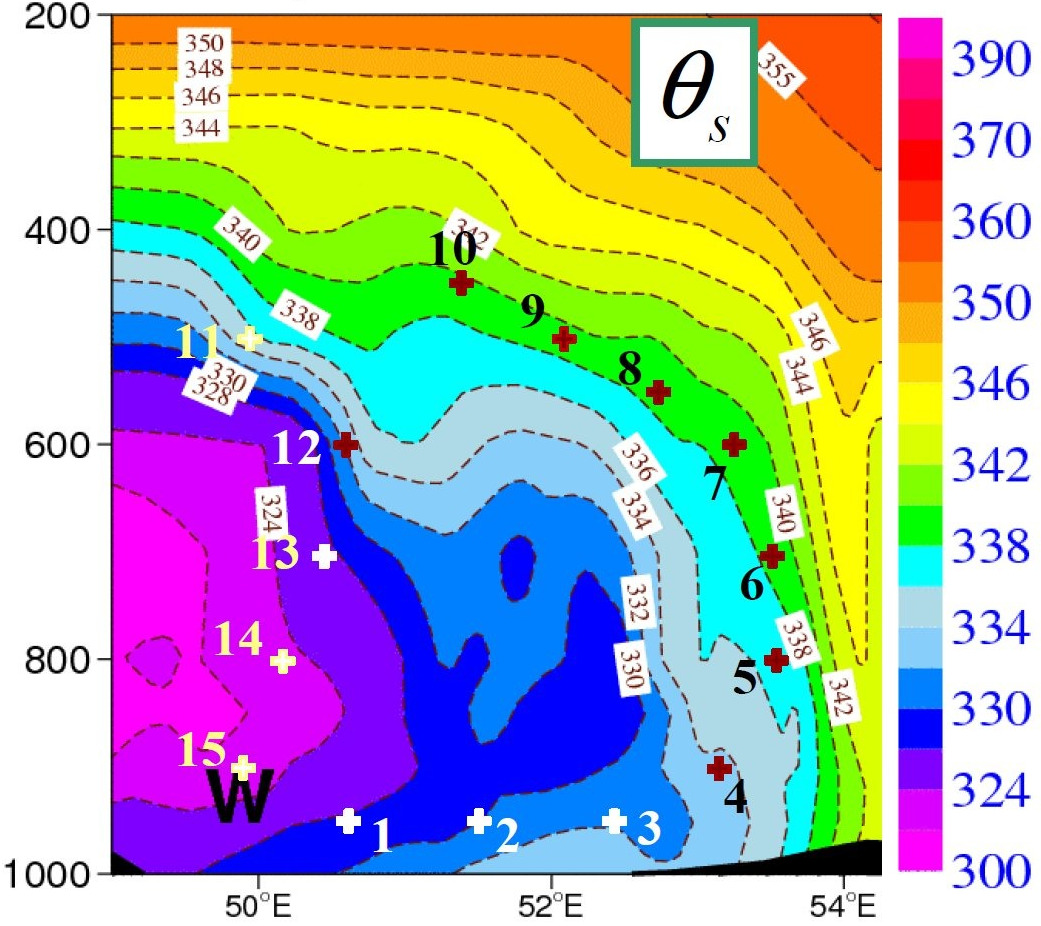}}
\vspace*{-3mm} 
\caption{\it The same as in Fig.~\ref{fig1}, but for the potential temperatures $\theta_{s/M11}$ in K computed by Eq.~(\ref{eq_thetas}).
} 
\label{fig4}
\end{figure}

The eyewall and the core of the hurricane are similar to Figs.~16 and 12 (top) plotted in \citet{Haw_Imb_MWR76} for the hurricane Inez, where $\theta_e$ was likely computed as an equivalent for $\theta'_w$.
The same high values of $\theta'_w$ observed in the eye of Inez are simulated in Fig.~\ref{fig1} for Dumile at the lower and upper levels.
The same area of minimum $\theta'_w$ and lower relative humidity is observed in the core in the $500$ to $700$ layer.

The cross section depicted in Fig.~\ref{fig4} for $\theta_s$ exhibits great differences in comparison with Fig.~\ref{fig1} valid for $\theta'_w$ and with Fig.~\ref{fig3} valid for $\theta_{e/B73}$:
there is a minimum of $\theta_s$ close to the surface in the core region; 
at a distance from the center $\theta_s$ exhibits a mid-tropospheric minimum at about $750$~hPa while it is about $1000$~m higher for $\theta'_w$ in Fig.~\ref{fig1}, for $\theta_{e/MPZ}$ in Fig.~2a of MPZ and for $\theta_{e/B73}$ in Fig.~\ref{fig3};
the isentropes computed and plotted with $\theta_s$ become tilted and almost horizontal above the level $600$~hPa in Fig.~\ref{fig4} while it is above the level $300$~hPa in Fig.~\ref{fig1} for $\theta'_w$ and in Fig.~\ref{fig3} for $\theta_{e/B73}$, or above $9000$~m with $\theta_{e/MPZ}$ in MPZ.
The isentropes computed with $\theta_s$ are therefore not compatible with the isolines of $\theta'_w$ or $\theta_{e/B73}$ and the more the relative humidity $H_l$ is large in Fig.~\ref{fig2} the more the isentropes differ.

Similarly, the aim of the following sections is to show that varying values for $r_v$ must have great impact on the definition and the plot of the isentropic surfaces considered in E86, E94 or MPZ.
To do so, all the moist-air equivalent potential temperatures and the entropies given by Eqs.~(\ref{eq_thetae_E94})-(\ref{eq_s_theta_es_E86_cpd}) are computed for a series of $15$ points selected arbitrarily and plotted in Figs.~\ref{fig1}-\ref{fig4}.
These $15$ points describe a sort of Carnot cycle inspired by the one described in \citet{Emanuel_86,Emanuel_91,Emanuel_ATMM_2004}.
The basic thermodynamic conditions ($p$, $T$, $r_v$, $q_l=q_i=0$) of these points are listed in Table~\ref{Table1}.
The dry descent follows a path of almost constant relative humidity between $60$ and $70$~\% (points $11$ to $15$), whereas the moist ascent follows a path of almost constant $\theta_s$ and moist-air entropy (points $6$ to $10$).

The impact of condensed water and of non-equilibrium terms is expected to be smaller than impact due to changes in $r_v$.
For the sake therefore of simplicity, the condensed water and the non-equilibrium effects are discarded (namely $T_l=T_i=T$ for the $15$ points, with $H_l=H_i=1$ and $q_l=q_i=0$ in the just-saturated regions).

\begin{table*}
\caption{\it The pressure (in units of hPa), temperature (in units of K), water-vapor mixing ratios (in units of g/kg) and relative humidity (over liquid water, in units of \%) for the $15$ points depicted in Figs.~\ref{fig1}-\ref{fig4}.
The potential temperature (in units of $K$) are 
the pseudo-adiabatic values ($\theta'_w$), 
the third-law based values $\theta_{s/M11}$ computed by Eq.~(\ref{eq_thetas}),
the saturating equivalent values $\theta_{es/E86}$ computed by Eq.~(\ref{eq_thetaes})
and 
the equivalent values $\theta_{e/MPZ}$ computed by Eq.~(\ref{eq_thetae_MPZ16}).
The moist-air entropies (in units of J~K${}^{-1}$~kg${}^{-1}$) are:  
   $s(\theta_{s/M11})$ computed by Eq.~(\ref{eq_s_thetas_M11})
with an offset of $-6850$~J~K${}^{-1}$~kg${}^{-1}$;
   $(s/q_d)(\theta_{es/E86})$ computed by Eq.~(\ref{eq_s_theta_es_E86_cpd}) 
with an offset of $-5650$~J~K${}^{-1}$~kg${}^{-1}$;
   $(s/q_d)(\theta_{e/MPZ})$ computed by Eq.~(\ref{eq_s_thetae_MPZ})
with an offset of $+550$~J~K${}^{-1}$~kg${}^{-1}$.
\label{Table1}
}
\vspace*{-1mm} 
\centering
\vspace*{2mm}
\begin{tabular}{|c|c|c|c|c|c|c|c|c|c|c|c|}
\hline
 N & $p$  &  $T$ &  $r_v$  &  $H_l$   & $\theta'_w$ & $\theta_{s/M11} $ & $\theta_{es/E86}$ & $\theta_{e/MPZ}$
   & \!\!$s(\theta_{s/M11})$\!\!  & \!\!$(s/q_d)(\theta_{es/E86})$\!\! &\!\!\!\! $(s/q_d)(\theta_{e/MPZ})$ \!\! \!\!\\ 
\hline
$1$  & $950$ & $295.10$ & $16.25$ & $91.9$ & $296.32$& $329.17$ & $345.36$ & $339.33$& $112.4$ & $222.0$ & $783.3$  \\ 
\hline
$2$  & $950$ & $296.56$ & $16.24$ & $84.1$ & $296.71$& $330.80$ & $351.29$ & $340.93$& $117.4$ & $239.2$ & $788.3$  \\ 
\hline
$3$  & $950$ & $296.12$ & $17.45$ & $92.6$ & $297.41$& $332.48$ & $349.34$ & $343.21$& $122.5$ & $233.6$ & $796.8$  \\ 
\hline
$4$  & $900$ & $294.07$ & $17.11$ & $97.5$ & $297.85$& $334.72$ & $348.84$ & $345.21$& $129.2$ & $232.1$ & $802.7$  \\ 
\hline
$5$  & $800$ & $290.16$ & $15.41$ & $99.9$ & $298.28$& $338.48$ & $350.30$ & $347.98$& $140.4$ & $236.3$ & $809.5$  \\ 
\hline
$6$  & $700$ & $285.08$ & $12.64$ & $100$  & $298.01$& $340.29$ & $350.15$ & $348.24$& $145.8$ & $235.9$ & $807.3$  \\ 
\hline
$7$  & $600$ & $278.05$ &  $8.94$ & $98.2$ & $296.95$& $339.74$ & $347.39$ & $345.57$& $144.2$ & $228.0$ & $795.3$  \\ 
\hline
$8$  & $550$ & $273.57$ &  $6.90$ & $95.9$ & $296.16$& $338.71$ & $345.23$ & $343.30$& $141.1$ & $221.7$ & $786.4$  \\ 
\hline
$9$  & $500$ & $270.03$ &  $4.87$ & $80$   & $295.58$& $339.52$ & $346.65$ & $342.81$& $143.5$ & $225.8$ & $782.9$  \\ 
\hline
$10$ & $450$ & $265.38$ &  $2.84$ & $60$   & $294.85$& $339.69$ & $346.88$ & $341.65$& $144.0$ & $226.5$ & $777.5$  \\ 
\hline
$11$ & $500$ & $268.89$ &  $3.35$ & $60.1$ & $293.90$& $335.01$ & $343.87$ & $337.33$& $130.1$ & $217.7$ & $765.0$  \\ 
\hline
$12$ & $600$ & $277.15$ &  $5.95$ & $70$   & $294.39$& $332.87$ & $345.19$ & $336.90$& $123.7$ & $221.6$ & $766.1$  \\ 
\hline
$13$ & $700$ & $282.52$ &  $7.40$ & $70$   & $293.45$& $327.41$ & $342.33$ & $332.41$& $107.0$ & $213.2$ & $753.5$  \\ 
\hline
$14$ & $800$ & $286.59$ &  $8.49$ & $70$   & $292.18$& $321.66$ & $338.42$ & $327.40$& $89.2$ & $201.7$ & $738.6$  \\ 
\hline
$15$ & $900$ & $292.28$ & $10.90$ & $70.1$ & $292.85$& $321.50$ & $342.54$ & $328.73$& $88.8$ & $213.8$ & $744.8$  \\ 
\hline
\end{tabular}
\vspace*{-4mm} 
\end{table*}



 \section{\large Impacts on potential temperatures} 
\label{section_impact_theta}
\vspace*{-4 mm}

\begin{figure}[h]
\centerline{\includegraphics[width=0.98\linewidth]{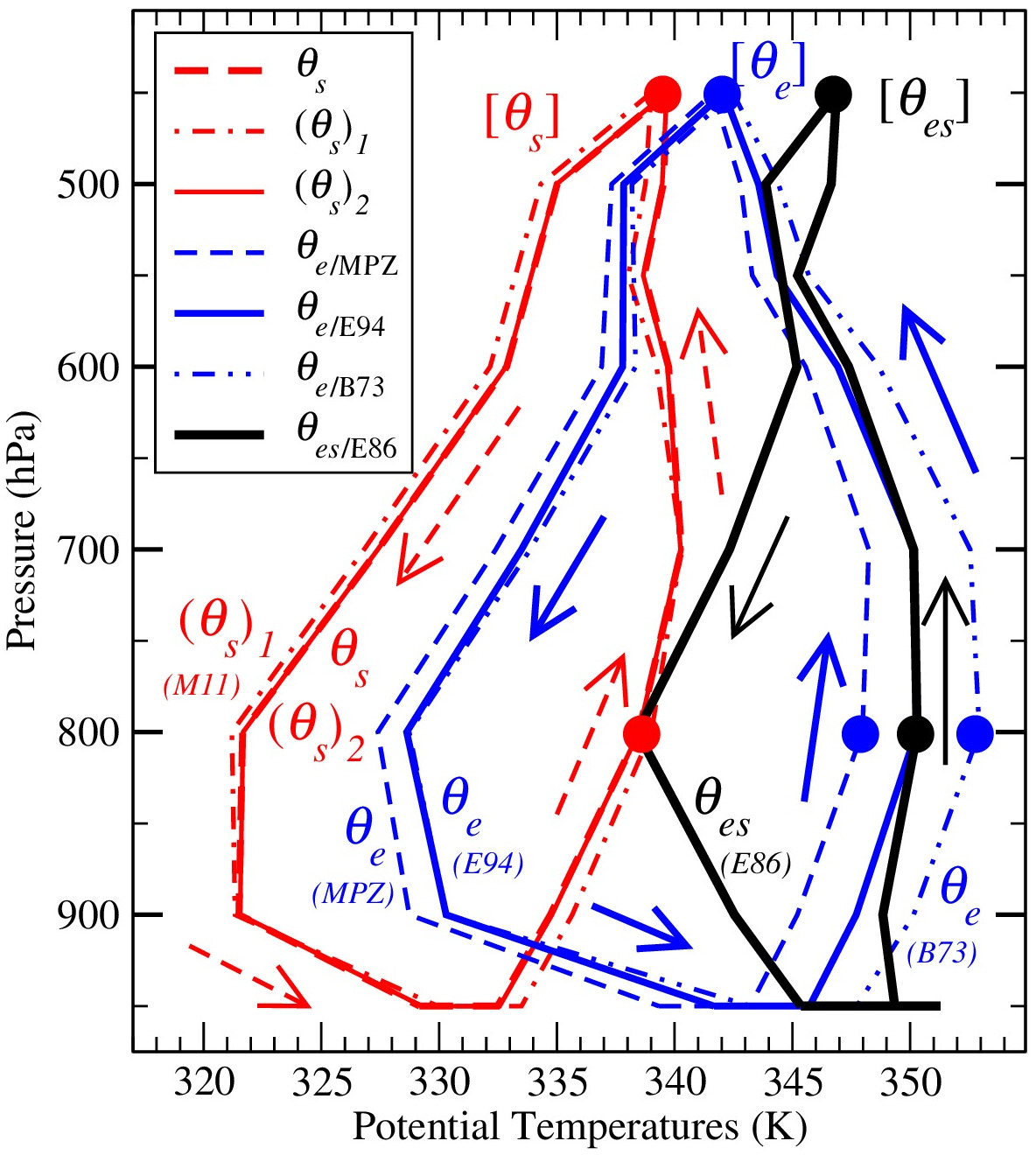}}
\vspace*{-3mm} 
\caption{\it The seven potential temperatures given by Eqs.~(\ref{eq_thetae_E94})-(\ref{eq_thetas2})  (in K) are plotted against the pressure (in hPa) for the $15$ points depicted in Figs.~\ref{fig1}-\ref{fig4}.
}
\label{fig5}
\end{figure}

All moist-air potential temperatures given by Eqs.~(\ref{eq_thetae_E94})-(\ref{eq_thetas2}) are plotted in Fig.~\ref{fig5}, with three of them listed in Table~\ref{Table1}.

As can clearly be seen, ${({\theta}_{s})}_1$, ${({\theta}_{s})}_2$ and ${\theta}_{s}$ remain close to each other with an accuracy of $\pm 0.8$~K for ${({\theta}_{s})}_1$, and with ${({\theta}_{s})}_2$ almost overlapping ${\theta}_{s}$.
This is proof that ${({\theta}_{s})}_1$ and ${({\theta}_{s})}_2$ are accurate increasing order approximations for ${\theta}_{s}$.

The equivalent formulations ${\theta}_{e/MPZ}$, ${\theta}_{e/E94}$ and ${\theta}_{e/B73}$ exhibit large discrepancies, especially in the warm and moist ascent where differences of $\pm 2.5$~K are observed between $950$ and $700$~hPa.
Moreover, the dry descent for the saturated version ${\theta}_{es/E86}$ is $10$~K warmer than that of other definitions of ${\theta}_{e}$.

The differences between the five formulations are small at high levels: they are less than $\pm 3$~K at $400$~hPa, because $r_v = 2.84$~g~kg${}^{-1}$ is small.
The impact is much larger at low level where  $r_v > 15$~g~kg${}^{-1}$, with $\theta_s$ about $15$~K colder than the ascent values of $\theta_e$ and $\theta_{es}$ at $950$~hPa.

``Isentropic'' surfaces or regions cannot therefore be the same if diagnosed through the use of values of either $\theta_s$, $\theta_e$ or $\theta_{es}$ and the comparisons described in MPZ which are based on analyses of altitude-$\theta_{e/MPZ}$ diagrams (MPZ's Figs.~4 and 6 to 9) might be invalid, since changes in the vertical may be of an opposite sign from the one for $\theta_s$.

Indeed, the solid disks plotted in Fig.~\ref{fig5} and the corresponding values in Table~\ref{Table1} show that ascents between the levels $800$ and $450$~hPa correspond to a small increase in $\theta_s$ of $+ 1.21$~K while they correspond to a large decrease in $\theta_{e/MPZ}$ of $- 6.33$~K.
Such a relative difference of about $7.5$~K is larger than the difference in equivalent potential temperatures of about $5$~K between the properties of updrafts and downdrafts described on page 1867 (right) in MPZ.

Another way in which to analyze the difference between $\theta_s$ and $\theta_{e/MPZ}$ or $\theta_{e/E94}$ is to consider the gap between descending and ascending values at $800$~hPa:  $\Delta \theta_s \approx 17 $~K and $\Delta \theta_{e/MPZ} \approx \Delta \theta_{e/E94} \approx 21$~K.
The difference is even larger for $\Delta \theta_{e/B73} \approx 24$~K.
On the other hand, the difference is much smaller for $\Delta \theta_{es/E86} \approx 12 $~K.

The important consequences of these findings is that changes in moist-air entropy represented by either $\theta_s$ or $\theta_{e/MPZ}$ cannot be simultaneously positive or negative, because otherwise it would be impossible to decide whether or not turbulence, convection or radiation processes would increase or decrease the moist-air entropy in the atmosphere.
Moreover, since isentropic processes or changes in entropy must be observable facts, it is impossible to consider that all definitions given by Eqs.~(\ref{eq_thetae_E94})-(\ref{eq_thetas2}) are equivalent: at most only one them can correspond to real atmospheric processes.

 \section{\large Impacts on moist-air entropies} 
\label{section_impact_s}
\vspace*{-4 mm}

\begin{figure}[h]
\centerline{\includegraphics[width=0.98\linewidth]{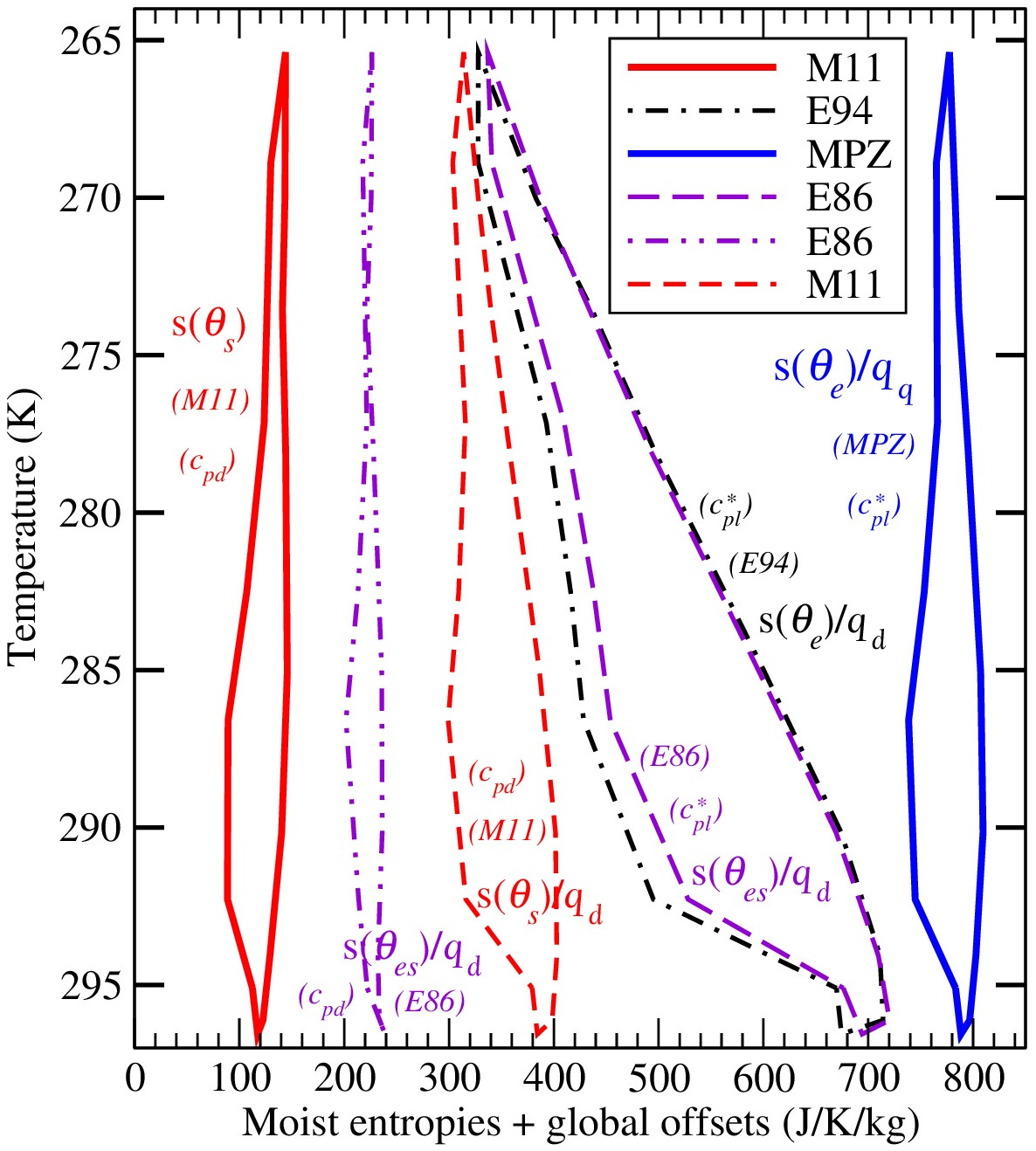}}
\vspace*{-3mm} 
\caption{\it The 6 moist-air entropies (in units of J~K${}^{-1}$~kg${}^{-1}$) defined by Eqs.~(\ref{eq_s_thetas_M11})-(\ref{eq_s_theta_es_E86_cpd}) plotted in a same temperature-entropy diagram.
Different global offsets are used for each entropy loop, in order to better separate and facilitate the comparison of the loops.
} 
\label{fig6}
\end{figure}

The issue of computing the relevant moist-air entropy is even harder than choosing one of the equivalent potential temperatures studied in the previous section, namely either $\theta_s$ or one of the versions of $\theta_e$ or $\theta_{es}$.
Since the aim of MPZ and \citet{Emanuel_86,Emanuel_91,Emanuel_ATMM_2004}, \citet{Pauluis_al_2010} or \citet{Pauluis_2011} is to analyze meteorological properties in moist-air isentropic coordinates, comparison of values of the moist-air entropy itself is needed.
Let us therefore plot in the temperature-entropy diagram depicted in Fig.~\ref{fig6} the values of the six moist-air entropies considered in section~\ref{section_s}.

The differences between the loops (Carnot cycles) are great.
Some of the loops are very narrow, whereas others are wide with a flared shape (a large gap between ascending and descending regions).
Some of the loops are almost vertical (with a small change of less than $\pm 20$~J~K${}^{-1}$~kg${}^{-1}$ in entropy between the surface and the upper air), whereas others exhibit a pronounced tilted feature (with a large decrease of $400$~J~K${}^{-1}$~kg${}^{-1}$ between the warm and moist regions and the cold and dry ones).

Again, since isentropic processes or changes in entropy must be observable facts, it is impossible to consider that all definitions given by Eqs.~(\ref{eq_s_thetas_M11}) to (\ref{eq_s_theta_es_E86_cpd}) are equivalent: at most only one of them can correspond to real atmospheric processes.

Another way in which to analyze the difference between the various formulations for moist-air entropy is to compute the heat input $W$ created by the closed loops in the temperature-entropy diagram, yielding
\begin{align}
 W & \: = \; \oint T(s) \; ds
\label{eq_W_int_T_ds}
\: .
\end{align}
Since $W$ is measured by the area of the loops in Fig.~\ref{fig6}, it is independent of the global offset chosen for each loop.

Values of $W$ listed in Table~\ref{Table2} show that it is impossible to consider that one can choose one formulation or another.
Since temperatures vary by about $30$~K and for an accuracy of about $0.2$~J~K${}^{-1}$~kg${}^{-1}$ for entropies, errors in computations of $W$ are about $6$~J~kg${}^{-1}$.
This means that the observed differences of the order of hundreds or thousands of J~kg${}^{-1}$ are significant.

More precisely, large factors of $3$ or $4$ are observed between $W$ computed with the third law value $s(\theta_s)$ and $W$ computed with $s(\theta_{es/E86})/q_d$ or $s(\theta_{e/E94})/q_d$.
Furthermore, the impact of $c^{\ast}_{pl}$ versus $c_{pd}$ is very large, leading to a factor of more than $6$ for the two heat inputs $W$ computed with the same potential temperature $\theta_{es/E86}$ ($452$ versus $2871$~J~kg${}^{-1}$).

\begin{table}
\caption{\it 
The heat input $W$ (in units of J~kg${}^{-1}$) given by Eq.~(\ref{eq_W_int_T_ds}) and computed for the six moist-air entropies considered in section~\ref{section_s}.
$W$ represents the areas of the six loops depicted in Fig.~\ref{fig6}.
The wind scale $\sqrt{2\,W}$ is in units of m~s${}^{-1}$).
\label{Table2}
}
\vspace*{-3mm} 
\centering
\vspace*{2mm}
\begin{tabular}{|c||c|c|c|c|c|c|c|}
\hline
``$\theta$'' & 
\!$\theta_{es/E86}\!\!$ & $\theta_{s}$ & \!$\theta_{e/MPZ}$\!\! & $\theta_{s}$ &\!\! $\theta_{es/E86}$\!\!  & \!\! $\theta_{e/E94}$\!\!  \\ 
\hline
``$s$'' 
& $s/q_d$ & $s$ & $s/q_d$ & $s/q_d$ & $s/q_d$ & $s/q_d$ \\ 
\hline
``$c_{p}$''
& $c_{pd}$ & $c_{pd}$  & $c^{\ast}_{pl}$ & $c_{pd}$ & $c^{\ast}_{pl}$ & $c^{\ast}_{pl}$ \\ 
\hline
 $W$ 
& $452$ & $892$ & $1213$ & $1572$ & $2871$ & $3564$\\ 
\hline
 $\!\!\sqrt{2\,W}\!\!$ 
& $30.1$ & $42.2$ & $49.2$ & $56.1$ & $75.8$ & $84.4$\\ 
\hline
\end{tabular}
\vspace*{-4mm} 
\end{table}

The impact of the definitions ``per unit of dry air'' versus the specific ones (namely ``per unit of moist air'') can be evaluated by comparing $s$ and $s/q_d$ for the same third law value $\theta_{s/M11}$.
The impact $1572 - 892 = 680$~J~kg${}^{-1}$ is great, leading to an increase of more than $75$~\% for the definition ``per unit of dry air''.

The impact of the choice of the different formulations for the moist-air entropy can be evaluated differently, by computing the wind scale $W = V^2/2$, with $V$ becoming a crude proxy for the surface wind that a perfect Carnot engine might produce.
The last line in Table~\ref{Table2} shows that $V$ would vary from about $30$ to more than $80$~m~s${}^{-1}$: this is unrealistic.


 \section{\large Conclusion} 
\label{conclusions}
\vspace*{-4 mm}

The isentropic analysis conducted in MPZ is likely a powerful tool for the investigation of moist-air energetics by the plotting of moist-air isentropes or by the computing of isentropic mass fluxes.
The quality and the realism of such an analysis relies on a clear definition of the moist-air entropy however.

It is shown in this Comment that the way in which the potential temperatures $\theta_s$, $\theta_e$ or $\theta_{es}$ are defined as ``equivalents'' of the moist-air entropy significantly impacts the computations and plots of isentropic surfaces, making the  ``isentropic'' analyses similar to the one published in MPZ uncertain.

It is shown moreover that the heat input computed for loops in the Carnot cycle is largely modified not only by the choice of $\theta_s$, $\theta_e$ or $\theta_{es}$, but also by the way in which the entropy itself is defined: $s$ or $s/q_d$; modified reference values for entropies; $c_{pd}$ or $c^{\ast}_{pl}$ in factor of the logarithm; other missing terms; {\it etc\/}

As for the issue associated with the vision ``per unit mass of dry air'', it can be understood as follows.
If isentropic processes are defined as in E94 and MPZ with constant values of $s/q_d$, one should modify accordingly the definitions of the geopotential, the wind components or the kinetic energy by plotting for instance $g\: z / q_d$, $u/q_d$, $v/q_d$ or $(u^2+v^2)/(2 \: q_d)$.
These definitions are unusual.
Moreover, if $s/q_d$ could be defined within a global constant $C$, the specific value $s$ would depend on $q_d \: C$, which varies with $q_d$ and renders the integral $S = \iiint s \: \rho \: d\tau$ indeterminate, because it would depend on $\iiint q_d \: C \: \rho \: d\tau$ where $C$ is an unknown term.

In conclusion, since moist-air isentropic surfaces are not subject to uncertainty in Nature, the third-law definitions $\theta_s$ and $s(\theta_s)$ given by Eqs.~(\ref{eq_thetas}) and (\ref{eq_s_thetas_M11}) are likely the more relevant, being as they are based on general thermodynamic principles and with specific values expressed per unit mass of moist air, as with all other variables in fluid dynamics.







\bibliographystyle{ametsoc2014}
\bibliography{arXiv_Comment_Mrowiec_JAS2016}

\end{document}